\documentclass{article}
\usepackage[utf8]{inputenc}
\usepackage{authblk}

\providecommand{\keywords}[1]
{
  \small	
  \textbf{\textit{Keywords---}} #1
}

\title {Einstein and the problem of confirmation by previously known evidence: A comment on Michel Janssen and Jürgen Renn's \emph{Einstein and the Perihelion Motion of Mercury}}
\author{Galina Weinstein}
\affil{\normalsize Reichman University, Herzliya, Israel. The Department of Philosophy, University of Haifa, Haifa.} 

\begin{document}

\maketitle

\begin{abstract}

In this paper, I comment on a recent paper by Michel Janssen and Jürgen Renn (\cite{Janssen}). In his published paper of November 18, 1915, Einstein presented a solution to the problem of the perihelion motion of Mercury and obtained the correct result of 43 seconds of arc per century. Before 1915, Einstein had established non-covariant field equations (the \emph{Entwurf} field equations). But in 1915, he changed his mind and dropped these equations and was led to general covariance. In a manuscript written in 1913, Einstein and his best friend Michele Besso tried to solve the \emph{Entwurf} field equations to attain the perihelion advance of Mercury. The end result arrived at by Einstein was $1821^{''}$.
To make a long story short, the field of a static Sun produced an advance of the perihelion of Mercury of $18^{''}$. Einstein kept quiet about this result and continued to work on the \emph{Entwurf} theory. Janssen and Renn ask: If we follow the prescriptions of Karl Popper, why did Einstein not accept that his \emph{Entwurf} theory had been falsified? They further ask: Given the undeniable importance of Besso's earlier calculations, why did Einstein not invite Besso as a co-author of his November 1915 paper on the perihelion motion of Mercury? Besso did not even get an acknowledgment in Einstein's paper (\cite{Janssen}, p. 7, p. 9).  
After a brief review of Einstein's and Besso's early attempts at calculating the advance of the perihelion of Mercury, I discuss falsification and confirmation by previously known evidence and then propose an answer to the above two questions. 

\end{abstract}


\section{Introduction}

In this paper, I comment on a recent paper by Michel Janssen and Jürgen Renn. Their paper discusses the \emph{Einstein-Besso manuscript} and Einstein's 1915 paper on the perihelion motion of Mercury (\cite {Janssen}). 
It is not the first time that Janssen has written on the \emph{Einstein-Besso manuscript} (for instance, \cite{Janssen1}, \cite{Janssen3}).
\vspace{1mm} 

I will first provide a short introduction to the most important steps in Einstein's route to his field equations. 
Starting in 1912, Einstein began collaborating with his schoolmate, Marcel Grossmann, who was a professor of mathematics in Zurich. Einstein started to work with tensors and jotted gravitational equations in a manuscript called the \emph{Zurich Notebook}, while Grossmann gradually updated him with new mathematical tools. 
In 1912, Einstein wrote a form of the Ricci tensor in terms of the Christoffel symbols and their derivatives. This was a fully covariant Ricci tensor in a form resulting from contraction of the Riemann tensor. He considered candidate field equations with a gravitational tensor constructed from the Ricci tensor. The gravitational tensor - called by scholars the "November tensor" - transforms as a tensor under unimodular transformations. It is called the November tensor because setting this tensor equal to the stress-energy tensor, multiplied by the gravitational constant, one arrives at the field equations of Einstein's first paper of November 4, 1915 (\cite{Einstein1}). In 1912, Einstein hoped he could extract the Newtonian limit from the November tensor. But he found it difficult to recognize that the November tensor reduces to the Newtonian limit. Thus, he finally chose non-covariant field equations. 

In March 1913, Einstein and Grossmann wrote a joint paper in which they established non-covariant field equations through energy-momentum considerations instead of the November tensor; these equations could also reduce to the Newtonian limit (\cite{Grossmann}). Like Einstein's general theory of relativity of November 1915, in the Einstein-Grossmann theory, the gravitational field was represented by a metric tensor and the action of gravity on other physical processes was represented in a form which remained unchanged under all coordinate transformations, i.e. by generally covariant equations. The only fly in the ointment was that the theory contained field equations that were not generally covariant.
This theory is called by scholars the Outline theory, and in German, the \emph{Entwurf} theory. Actually, Einstein also used this name (Einstein to Lorentz, August 14, 1913, \cite{CPAE5}, Doc. 467). 

Since the Mercury anomaly could not be easily explained in the framework of Newtonian gravitation theory, it presented a good opportunity to theoretically check the \emph{Entwurf} theory of gravitation. So, in June 1913, Einstein's closest friend and conﬁdant, the engineer Michele Besso, came to visit him in Zurich, and they both tried to solve the new \emph{Entwurf} field equations of Einstein to attain the perihelion advance of Mercury in the field of the static Sun. Besso was much more than just an engineer. In 1926 Einstein wrote that Besso had extraordinary knowledge in pure science (Einstein to Zangger, December 21, 1926 \cite{CPAE15}, Doc. 436). Besso had a deep knowledge of physics and mathematics. He visited Einstein again in August 1913. During this time, Einstein and Besso continued to work on the project. Their joint work consists the \emph{Einstein-Besso manuscript}, which contains calculations with little explanations.

During 1913-1914, Einstein elaborated the \emph{Entwurf} theory to the point that its field equations were valid for adapted coordinate systems and he proposed a few versions of the hole argument against general covariance. 

Around October-November 1915, Einstein changed his mind and dropped the \emph{Entwurf} field equations. He postulated that his gravitational tensor should be invariant under unimodular transformations. This finally led him to general covariance. The long and devious route Einstein took from 1912 until the above step finally led him back to his starting point, namely to the November tensor of 1912. Einstein gradually expanded the range of the covariance of his gravitational field equations and on November 18, 1915 he presented a solution of the problem of the perihelion motion of Mercury, i.e. he obtained the correct result of 43 seconds of arc per century. All in all Einstein published a series of four papers in November 1915 (see \cite{Weinstein}, chapter D).
\vspace{1mm} 

For now, let us concentrate on the \emph{Einstein-Besso manuscript} and Einstein's November 18, 1915 paper on the perihelion motion of Mercury. Janssen writes that in 1988, the Einstein Papers Project obtained a copy of the \emph{Einstein-Besso manuscript} in possession of descendants of Besso. In the 1970s, Pierre Speziali edited the Einstein-Besso correspondence (\cite{Besso}) and became friendly with Besso's son Vero who gave him the manuscript. Speziali approached the director of the Einstein Papers Project and made a copy of the manuscript available to the project. A portion of the manuscript (which consists of 53 pages, half of them in Einstein’s hand and the other half in Besso’s, but there might well be more pages) was published in the \emph{Collected Papers of Albert Einstein} (\cite{CPAE4}). Historians of science then began to analyze the manuscript (\cite{Janssen3}). 
After discovering the \emph{Einstein-Besso manuscript}, scholars realized that (\cite{CPAE4}):

1) Einstein had already calculated the perihelion motion of Mercury in 1913 in collaboration with Besso.

2) The method used in the \emph{Einstein-Besso manuscript} is virtually identical to the method Einstein used in his November 18, 1915 paper on the perihelion motion of Mercury.
\vspace{1mm} 

In their new paper, Janssen and Renn argue that if Einstein had followed the prescriptions of Karl Popper, he should have accepted that his \emph{Entwurf} theory had been falsified. But during 1913-1914, Einstein did nothing of the sort. According to the \emph{Entwurf} theory, the field of a static Sun produced an advance of the perihelion of Mercury of $18^{''}$ (18 seconds of arc per century). Einstein kept quiet about this result and continued to work on the \emph{Entwurf} theory. Of course, the story is not so simple and the end result which was given in the \emph{Einstein-Besso manuscript} by Einstein was: $1821^{''}$ and it was corrected. This is discussed in section \ref{1}.

Janssen and Renn further claim that given the undeniable importance of Besso's earlier calculations, "one can legitimately ask why Einstein did not invite Besso as a co-author" of his November 18, 1915 paper on the perihelion motion of mercury. "As it happened", say Janssen and Renn, "Besso did not even get an acknowledgment" in Einstein's November 18, 1915 paper (\cite{Janssen}, p. 7). 

After a brief review of Einstein's and Besso's early attempts at calculating the advance of the perihelion of Mercury, I discuss falsification (section \ref{1}) and confirmation (section \ref{2}) of a theory by evidence. A problem in confirmation theory which was mainly considered by Clark Glymour, and which has been the subject of many papers, is confirmation by old evidence. Einstein derived out of general relativity the perihelion motion of Mercury discovered more than half a century earlier. I end section \ref{2} with a discussion of confirmation of a theory by previously known evidence. 

In sections \ref{2} and \ref{3} I propose an answer to two questions: 

1) Why did Einstein not reject his \emph{Entwurf} field equations in the face of the wrong result $18^{''}$?  

2) Why was Besso not invited as a co-author of the 1915 paper on the perihelion motion of Mercury, even though he contributed to the calculations in the \emph{Einstein-Besso manuscript}?  

\section{Falsification} \label{1}

In the summer of 1913, in the \emph{Einstein-Besso manuscript}, Besso and Einstein tried to solve the \emph{Entwurf} vacuum field equations and calculate the perihelion advance of Mercury in the field of the static Sun. 
According to the \emph{Entwurf} theory, the field of a static Sun produces an advance of the perihelion of Mercury of (\cite{CPAE4}, p. 351): 

\begin{equation} \label{equation 1}
1.25 \pi \frac{A}{a\left (1-e^2\right)} \vspace{1mm} 
\end{equation}

\noindent per revolution, where $A$ is proportional to the gravitational constant and the mass of the Sun, $a$ represents the semi-major axis and $e$ - the eccentricity of the elliptical orbit. 

According to the \emph{Entwurf} theory, the field of a static Sun produces an advance of the perihelion of Mercury of $18^{''}$ but the end result which is given by Einstein is: $1821^{''}$. Next to this result Einstein wrote in the \emph{Einstein-Besso manuscript}: “independently checked” (\cite{CPAE4}, p. 28). 
Einstein's result was a factor $100$ too large. The Sun, which in Newton's theory produces no perihelion motion at all, in the \emph{Entwurf} theory, produces a perihelion motion of more than three times the size of the total perihelion motion that is observed. Fortunately, Einstein and Besso found a mistake in the numerical calculation – an error factor of 10 for the mass of the Sun. Since the perihelion motion is proportional to the square of this quantity, the final result was an error factor of 100. 
Several pages ahead there is a correction in Besso's hand of the erroneous value for the mass of the Sun that Einstein had used. On another page, Einstein himself corrected the old value for the mass of the Sun.\footnote{More specifically, on page 30 of the \emph{Einstein-Besso manuscript} another attempt is made by Einstein to find the perihelion advance of Mercury in the field of a static sun, because Einstein found a mistake in the previous calculation regarding the evaluation of the mass of the Sun. He wrote the new value as: $M = 1.96 \times 10^{33}$. But the new calculation contains several other errors. The end result of this calculation, in fractions of $\pi$ per half a revolution, is a perihelion advance of: $1.65 \times 10^{-8}$. Below this number Einstein wrote another number: $3.4 \times 10^{-8}$. This is probably the number that Einstein expected to obtain with the new value of the mass of the Sun. After correcting the errors in the calculation this is indeed the number one obtains and this gives an advance of the perihelion of Mercury of $18^{''}$ per century (\cite{CPAE4}).}

Janssen and Renn explain that Einstein found the precession per half a revolution in seconds of arc and the precession in 100 years in seconds. "The correct result of $18^{''}$ is not stated anywhere in the manuscript, although, [...] there are clear indications that Einstein and Besso 
discovered the source of the erroneous factor 100 in the result stated on" p. 28. It seems that Besso was the first one who found the source of the error, say Janssen and Renn (\cite{Janssen}, p. 7, p. 46).

Be that as it may, $1821^{''}$ is erroneous because of Einstein's computation error and $18^{''}$ is wrong because of Einstein's \emph{Entwurf} theory. Janssen and Renn conclude: "The theory would have been decisively refuted had it truly predicted $1800^{''}$; that it could only account for less than half the missing seconds was undoubtedly disappointing but not particularly troublesome" (\cite{Janssen}, p. 6). 
The point is that the \emph{Entwurf} hypothesis of non-covariance is refuted by the evidence, i.e. $18^{''}$. Yet Einstein never mentioned the above disappointing result until he recalculated it on the basis of the correct November 1915 theory. 
\vspace{1mm} 

So, let us try to answer the question: 

\noindent Why did Einstein not reject his \emph{Entwurf} field equations in the face of the wrong result $18^{''}$? 

I shall now try to answer this question starting with rotation and ending with falsification.   

On pages 41-42 of the \emph{Einstein-Besso manuscript}, Einstein checked whether the Minkowski metric in rotating coordinates was a solution of his \emph{Entwurf} vacuum field equations. 
To find this, he used the same approximation procedure he had used to calculate the field of the Sun for the perihelion advance of Mercury. He first substituted the components of the rotation metric into the \emph{Entwurf} vacuum field equations in a first-order approximation. He thus checked whether the metric field of the rotating system was a solution of the field equations of the \emph{Entwurf} theory. However, in substituting the first-order components of the rotation metric into the \emph{Entwurf} vacuum field equation, Einstein made mistakes. As a result of his mistakes he obtained the answer he wanted. He asked whether the solution for the (44-component of the metric tensor) $g_{44}$ found from the \emph{Entwurf} field equations was the same as the $g_{44}$ one would obtain through transformation of the Minkowski metric from an inertial system to the rotating system. Einstein's answer in the \emph{Einstein-Besso manuscript} was yes, and next to the solution he wrote that it was correct (\cite{CPAE4}). 

Janssen and Renn note that Einstein made several mistakes in the above calculation. He incorrectly read off the components of the metric from the line element and wrote a plus sign which should have been a minus sign, and thus made a sign error. When these mistakes are corrected, the metric field of the rotating system is not a solution of the field equations of the \emph{Entwurf} theory (\cite{Janssen}, p. 47).

The point is that, Einstein related between the problem of rotation and the perihelion of Mercury. For small angular velocities, the metric field of a rotating system has the same general form as the metric field of the Sun – considered by Einstein and Besso in their 1913 manuscript for calculating the perihelion of Mercury. Thus, Einstein could use the approximation procedure used to find the metric field of the Sun to second-order to find the metric in a rotating coordinate system. He could also do the opposite; refer from his calculation of the rotating coordinate system about the case of the metric field of the Sun and Mercury's perihelion. 

The \emph{Einstein-Besso manuscript} remained with Einstein in Zurich and in early 1914, he sent it to Besso, urging his friend to finish their project. Besso added more calculations. It seems that in 1913 Einstein "did not follow the prescriptions of Karl Popper" \cite{Popper} or Carl Hempel (Hempel's version of falsification) \cite{Hempel} because he was under the impression that the Minkowski metric in rotating coordinates was a solution of his \emph{Entwurf} vacuum field equations. Since this problem is intimately related to the perihelion advance of Mercury, it is reasonable to assume that Einstein hoped the $18^{''}$ value would eventually be corrected by adding other terms.

In 1913, Gunnar Nordström constructed an alternative theory to Einstein's \emph{Entwurf} theory. Nordström's theory was simpler than the one suggested by Einstein and more related to special relativity.\footnote{Nordström left the speed of light constant and adapted his theory of gravitation to the special principle of relativity in such a way that gravitational and inertial masses were equal. He thus accepted a version of Einstein's equality of inertial and gravitational masses and included it in his theory. Unlike Einstein's tensor theory, Nordström proposed a scalar theory.} Nordström's and Einstein's theories had almost the same empirical consequences. Both Nordström's and Einstein's theories predicted a red shift of spectral lines. But according to Nordström's theory, there was no bending of light rays in a gravitational field.  In fact, the \emph{Entwurf} theory predicted a value for the deflection of light in a gravitational field of the Sun, 0.83 arc seconds, which much later turned out to be wrong. On the other hand, Einstein's 1915 generally covariant theory yields twice that value, 1.7 arc seconds. But Einstein had no way to know that the value 0.83 was wrong.  
The fact of the matter is that neither theories could yield the correct value of the perihelion motion of Mercury. On page 53 of the \emph{Einstein-Besso manuscript} (the calculations added by Besso in 1914), Besso calculated the perihelion motion of Mercury using Nordström's theory, and the strategy followed in these calculations was the same as that followed by Einstein and Besso when they had both solved the \emph{Entwurf} field equations in 1913 to find the perihelion advance for Mercury of 18 seconds of arc per century (\cite{CPAE4}). 

Thus, in 1913 there were two incompatible theories, each could not account for the motion of the perihelion of Mercury. Before 1915, the \emph{Entwurf} theory remained without empirical support and a decision in favor of one or the other theory – the \emph{Entwurf} or Nordström's – was impossible on empirical grounds.
But from Einstein's point of view, Nordström's theory had a big flaw: it did not satisfy Ernst Mach's ideas, i.e. it did not account for rotation. As we have already seen, neither did Einstein's \emph{Entwurf} theory explain rotation. 
\vspace{1mm} 

Upon revisiting his calculations in September-October 1915, Einstein was willing to accept his mistakes.
\vspace{1mm} 

First, already in an exchange of letters and postcards that began in March 1915 and ended in May 1915, Tulio Levi-Civita presented his objections to Einstein's adapted coordinate systems. Einstein did not agree with Levi-Civita and stubbornly clung to what had remained from his \emph{Entwurf} theory. There was even a moment in which Einstein regained full confidence in his \emph{Entwurf} theory and he admitted to Levi-Civita, through the deeper considerations to which the latter interesting letters have led him, that he became even more firmly convinced that his theory was correct (see \cite{Weinstein}, Chapter D).  

In the autumn of 1915, Einstein eventually accepted that he had chosen a problematic Lagrangian function. Einstein's 1914 ten field equations determine the ten functions $g^{\mu \nu}$ but the $g^{\mu \nu}$ must also satisfy four conditions for the special choice of adapted coordinate systems. In 1914, he thought that the selection of the above Lagrangian could be supported by this restriction and he provided a proof that, the above Lagrangian is uniquely selected by the requirement that it is invariant for adapted coordinate systems. Einstein believed that the variation of the Lagrangian uniquely led to the \emph{Entwurf} field equations. Therefore, he thought he had demonstrated that his \emph{Entwurf} field equations were the only equations that were invariant for adapted coordinate systems. 
Around October-November 1915, he realized that his Lagrangian function did not uniquely lead to the \emph{Entwurf} field equations that were invariant for adapted coordinate systems. Consequently, covariance with respect to adapted coordinate systems was a wrong path and Einstein understood that he had to require general covariance.
\vspace{1mm} 

Second, Janssen and Renn argue that in 1913 Besso discussed with Einstein rotation and the possibility that the Minkowski metric in rotating coordinates was not a solution of his \emph{Entwurf} field equations. Einstein though would nonetheless continue to cling to his \emph{Entwurf} theory (I will further elaborate on this suggestion later in this section).  

In the autumn of 1915, Einstein finally recognized that the metric field describing a rotating system was not a solution of the \emph{Entwurf} vacuum field equations. Einstein corrected his initial mistakes and found that his field equations were not satisfied and the metric field describing a rotating system was not a solution of these equations. 
He obtained a solution for the $g_{44}$ from the \emph{Entwurf} field equations and demonstrated that it was not the same as the one obtained through direct transformation of the Minkowski metric to the rotating coordinate system. 

On September 30, 1915 Einstein sent a letter to Erwin Freundlich saying he had found a blatant contradiction in his theory of gravitation. He demonstrated to Freundlich that the rotation metric was not a solution of his \emph{Entwurf} field equations. He realized that he must have done a calculation error somewhere in his work. The above realization seemed to have stimulated what Einstein had already discovered earlier that, Mercury's perihelion motion was too small, namely $18^{''}$. Indeed, Einstein told Freundlich that he had no doubt that his solution for the perihelion motion of Mercury (from 1913) was also suffering from the same problem. He therefore immediately related between the problem of rotation and the perihelion motion of Mercury (Einstein to Freundlich, September 30, 1915, \cite{CPAE8}, Doc. 123). 
But recall that in the \emph{Einstein-Besso manuscript} Einstein had already connected the two problems.

The day afterwards, Einstein sent a letter to Otto Naumann, in which he redid the calculation of pages 41-42 of the \emph{Einstein-Besso manuscript} and found that the metric field describing a rotating system was not a solution of the \emph{Entwurf} field equations. He also demonstrated that the solution obtained from the \emph{Entwurf} field equations, was not the same as the one obtained through direct transformation of the Minkowski metric to the rotating coordinate system (Einstein to Naumann, October 1, 1915, \cite{CPAE8}, Doc 124, note 5). 
This indicated to Einstein that his solution for the perihelion motion of Mercury suffered from the same fault as the solution for a rotating coordinate system.

Janssen writes: "The story suggests that it was the problem with rotation rather than the other two problems [...] that dealt the decisive blow to the \emph{Entwurf} theory" (\cite{Janssen4}, p. 127). It seems to me that for Einstein, the two problems - that of the perihelion advance of Mercury and that of rotation - were interconnected. But what led Einstein to give up his \emph{Entwurf} theory in 1915 was first and foremost the entanglement between rotation and Mercury. In 1915 Einstein redid the calculation of pages 41-42 of the \emph{Einstein-Besso manuscript} and discovered his arithmetic error and thus, found out by himself that the metric field describing a rotating system was not a solution of the \emph{Entwurf} field equations. He then related this problem to Mercury's perihelion.    

Now let us return back to 1913. It appears that the reason Einstein did not accept that his \emph{Entwurf} theory was falsified, even though it predicted the wrong result for the perihelion motion of Mercury, was that he related between the problem of rotation and that of the perihelion motion of Mercury. Einstein's calculations in the \emph{Einstein-Besso manuscript} concerning rotation involve the same approximation procedure Einstein had used to calculate the field of the Sun for the perihelion advance of Mercury. 

The point is that in his 1935 book, \emph{Logik der Forschung} \cite{Popper}, Popper proposes an absolute notion of falsification which demands a refutation of a theory by a single individual counter instance. 
Although the result Einstein and Besso had obtained $18^{''}$ was a single counter example to Einstein's \emph{Entwurf} theory, in the \emph{Einstein-Besso manuscript} Einstein and Besso nonetheless tried to resort to additional effects that might contribute to the perihelion advance of Mercury. But these effects did not produce the desired value of $43^{''}$ (\cite{CPAE4}). Nevertheless, Einstein was not bothered of the small perihelion advance of Mercury $18^{''}$, as long as he believed that the rotating metric was a solution of his \emph{Entwurf} field equations. The reason is that from his point of view, these problems were entangled. 

Janssen and Renn argue that, despite warnings recorded by Besso in his memo - notes that Besso wrote of discussions with Einstein during their attempt to account for the perihelion anomaly of Mercury on the basis of the \emph{Entwurf} theory - Einstein was nonetheless firmly convinced that the rotation metric was a solution of the \emph{Entwurf} field equations (\cite{Janssen}, p. 47). 

Besso wrote on the first page of his notes of discussions with Einstein, called \emph{Besso's memo}: "One can therefore not think of rotational forces as produced by the rotation of the fixed stars according to the Einsteinian gravitational equations". 
In the top-right corner of the first page of the memo,
Besso wrote: “28 VIII 13.” (\cite{Janssen2}, p. 806, p. 787). 
Hence, Besso dated the first page of the memo to be from August, 28, 1913. 
Accordingly, Janssen and Renn interpret this as indicating that Besso told Einstein that the rotation metric was not a solution of the \emph{Entwurf} field equations. Einstein though would often stubbornly stick to his errors and mistakes and he was unwilling to listen to warnings that people tried to bring to his attention. 

But we know that in August, 1913 – on the second time Besso visited Einstein in Zurich – Einstein was dissatisfied with his \emph{Entwurf} field equations. 
On August 14, 1913, he wrote to Hendrik Antoon Lorentz that the problem hooked him so much, that his confidence in the admissibility of the \emph{Entwurf} theory was still shaky. Einstein was disturbed by the thought of the non-general covariance of his gravitational field equations. If his field equations permitted only linear transformations and were not generally covariant, then Einstein felt the theory refuted its foundations and it thus stood unsupported with no basis. Two days later, however, Einstein wrote to Lorentz that he had found a way out of this muddle. Using the law of momentum and energy conservation, Einstein limited the choice of reference systems to certain allowed ones. With respect to these reference systems, the law of momentum and energy conservation holds and the general linear transformations remain as the only possible choice. Einstein was most satisfied and he told Lorentz that now the theory gave him pleasure after he managed to eliminate this ugly dark spot (Einstein to Lorentz, August 14, 16, 1913, \cite{CPAE5}, Doc. 467, 470).

This could hint that for a very short while, Einstein could have thought that his \emph{Entwurf} theory was falsified. Besso had probably pinpointed the problem of rotation. He was the one that put his hand on the problem. It is then reasonable to assume that in August 1913, rotation was at least one of the reasons why \emph{for a brief moment} Einstein was dissatisfied with his \emph{Entwurf} theory. Another reason could be Mercury's perihelion. But very soon afterwards, Einstein fell straight into the trap of the hole argument against general covariance. 
Two years later, in January 1915, Einstein told Lorentz that an argument he presented him has already been stressed by Ernst Mach. "But it is best illustrated by a comparison that my friend Besso had thought up last year" (Einstein to Lorentz, January 23, 1915, \cite{CPAE8}, Doc. 47). It seems that at that time, Einstein tried to find a solution to the problem that his rotation metric was not really a solution of his \emph{Entwurf} field equations. Doing so enabled him to cling even more tightly to his hole argument and his \emph{Entwurf} theory.

There is still the possibility that before 1915, Einstein refused to accept that the rotating metric was not a solution of his \emph{Entwurf} field equations. And only in 1915, when he redid the calculation of pages 41-42 of the \emph{Einstein-Besso manuscript}, did he discover his arithmetic error and found out by himself that the metric field describing a rotating system was not a solution of the \emph{Entwurf} field equations. This scenario seems to me the most reasonable one.     

After 1913, Einstein's belief in the \emph{Entwurf} theory was strengthened by deriving the Newtonian law of gravitation from the linear approximation of the \emph{Entwurf} field equations for the static case (for which the spatial metric was flat) and by developing the hole argument. After all, a year earlier, Einstein failed to extract the Newtonian limit from the November tensor and his calculations entangled the Newtonian limit and conservation of momentum-energy. On the other hand, the \emph{Entwurf} field equations satisfied both the Newtonian limit and conservation of momentum-energy and so, it was difficult to free Einstein from the clutches of the hole argument and later from his adapted coordinate systems (see \cite{Weinstein}, Chapter D).

\section{Confirmation} \label{2}

As said in the previous section, sometime in October 1915 Einstein dropped the \emph{Entwurf} theory. Starting on November 4, 1915, he gradually expanded the range of the covariance of his field equations \cite{Einstein1}. On November 11, 1915, Einstein wrote the field equations of gravitation in a generally covariant form. In the addendum of November 11, 1915, he wrote these equations with the source term \cite{Einstein2} and on November 18, 1915, the equations were vacuum field equations \cite{Einstein3}. 
In his November 18 paper \cite{Einstein3}, Einstein solved the problem of the perihelion of Mercury. He transferred the basic framework of the calculation from the \emph{Einstein-Besso manuscript} and corrected it according to his November 18, 1915 generally covariant vacuum field equations and geodesic equation written in terms of the Christoffel symbols.
Einstein used these vacuum field equations for calculation through successive approximations of the gravitational field of the Sun. He calculated the metric field of the Sun using these vacuum field equations in a first-order approximation, substituted the result of this calculation back into the vacuum field equations, obtained a second-order approximation and solved the vacuum field equations to obtain more accurate approximations for the metric field of the Sun. 

Einstein's solution was an ellipse that had a major axis that was not constant and rotated. The field of a static Sun produced a precession of the perihelion, an advance of the perihelion of $43$ seconds of arc per century (\cite{Einstein3}, p. 838):

\begin{equation} \label{equation 2}
3 \pi \frac{A}{a\left (1-e^2\right)}
\end{equation}

\noindent per revolution.

Compare equation (\ref{equation 2}) with equation (\ref{equation 1}). The $1.25 \pi$ equation leads to an advance of the perihelion of $18^{''}$. Multiply this by 2.4 and you get the advance of the perihelion of $43^{''}$ and the $3 \pi$ equation.

In the \emph{Entwurf} theory, the components of the gravitational field were expressed as the gradient of the metric tensor. On the other hand, in November 1915, Einstein expressed the components of the gravitational field in terms of the Christoffel symbols of the second kind. So, in 1915, the components of the gravitational field of the static Sun were the Christoffel symbols. 

Einstein then wrote equations of motion for a point mass moving in the gravitational field of the Sun. A planet in free fall in the gravitational field of the Sun moves on a geodesic line according to the geodesic equation. The point mass moves on a geodesic line under the influence of the gravitational field of the Sun. Einstein calculated the equations of the geodesic lines in this space and compared them with the Newtonian equations of the orbits of the planets in the solar system. 
\vspace{1mm} 

Einstein discovered two errors in the 1913 calculations:
\vspace{1mm} 

1) In the \emph{Einstein-Besso manuscript}, Einstein calculated the metric field of the Sun using the \emph{Entwurf} field equations in a first-order approximation, and found that the static spatial metric was \emph{flat}. Thus to first-order, Mercury moves along a geodesic curve that depends on \emph{flat} components of the metric. In November 1915, when Einstein redid this calculation, he calculated the metric field of the Sun using the November 18, vacuum field equations of the general theory of relativity in a first-order approximation. Einstein wrote to Besso that he was surprised by the occurrence of \emph{non-flat} components of the first-order (static spatial) metric and he discovered that Mercury moves along a geodesic curve that depends on \emph{non-flat} components of the metric (Einstein to Besso, December 10, 1915, \cite{CPAE8}, Doc. 162; \cite{Stachel}, pp. 304-306). Unlike the \emph{Entwurf} field equations, the 1915 ones are generally covariant and a material point in a gravitational field moves on a geodesic line where both equations are written in terms of the Christoffel symbols. 
\vspace{1mm} 

2) Besso managed to write his "area law" in the form of the Newtonian angular momentum conservation, but Kepler's second law (the radius vector from the sun to the planet sweeps out equal areas in equal times) did not hold for the \emph{Entwurf} theory. On the other hand, in his November 18, 1915 paper, Einstein emphasized that Kepler's laws hold exactly for his generally covariant general theory of relativity. Hence, Einstein corrected and adjusted Besso's derivation of 1913 according to his generally covariant theory. He then stressed in his November 18, 1915 paper that the equations that determine the motion of the planet for the case of circular motion give no deviation from Kepler's laws (\cite{CPAE4};\cite{Einstein3}, p. 837). 
\vspace{1mm} 

Einstein attempted to obtain a solution, without considering whether or not there was only one possible unique solution, exactly as he had done with Besso in 1913. 
Einstein's theory was consistent with the evidence. Moreover, unlike the \emph{Einstein-Besso manuscript}, the result $43^{''}$ confirmed Einstein's general theory of relativity of November 18, 1915 and it was accepted as an account for the advance of the perihelion of Mercury. 

On the very day Einstein submitted his paper on the perihelion motion of Mercury to the \emph{Proceedings of the Royal Prussian Academy of Sciences, Berlin}, he wrote to David Hilbert. He told him that he was going to submit "today" a paper in which he derived quantitatively out of general relativity, without any guiding hypotheses, the perihelion motion of Mercury, "discovered by Le Verrier" \emph{more than half a century earlier}.
Einstein added that this was not achieved until then by any gravitational theory (Einstein to Hilbert, November 18, 1915, \cite{CPAE8}, Doc. 148). 
\vspace{1mm} 

From the above letter we learn two things: 
\vspace{1mm} 

\noindent \emph{First}, Einstein explained to Hilbert that his general relativity (the new November 18, 1915 general relativity) entails ("without any guiding hypotheses") the advance of the perihelion of Mercury and therefore, it has to be considered as confirming this evidence (see definition in \cite{Hempel}, p. 102). 
\vspace{1mm} 

\noindent \emph{Second}, the precession of the perihelion of Mercury was, an unexplained phenomenon that bothered scientists even before Einstein had advanced his general theory of relativity. 
If we take into consideration Hempel's logical confirmation theory, then unlike Newtonian classical gravitational theory, the result is evidence in favour of the general theory of relativity and so, as said above, the correct value of the anomaly confirms Einstein's 1915 general theory of relativity. But from the point of view of the Bayesian confirmation theory, there seems to be a problem here. 
\vspace{1mm} 

In the chapter, “Why I am not a Bayesian”, Glymour asks: “When does a piece of evidence confirm a hypothesis according to the Bayesian scheme of things? The natural answer is that it does so when the posterior probability of the hypothesis is greater than its prior probability […]”.  A Bayesian learns new facts and each time he learns a new fact he revises his degrees of belief by conditionalizing on the new fact. The discovery that the fact is the case has confirmed those hypotheses whose probability after the discovery is higher than their probability before (\cite{Glymour2}, p. 38, \cite{Glymour1}, p. 83). 
Glymour, however, thinks this account is unsatisfactory for a few reasons the most problematic of which is the following: “Scientists commonly argue for their theories from evidence known long before the theories were introduced […] Newton argued for universal gravitation using Kepler's second and third laws, established before the Principia was published” (\cite{Glymour1}, pp. 85-86). 

According to Glymour, Einstein argued for his theory of relativity from evidence known long before his theory was introduced. Glymour gives the example of the perihelion of Mercury: in 1915 Einstein derived by general relativity the anomalous advance of Mercury's perihelion discovered more than half a century earlier. 
Glymour, however, argues that if one attempts to explain old evidence, one encounters a problem. He shows that old evidence can in fact confirm new theory, but according to Bayesian conditionalization it cannot. 

By the same token, one might argue that in his 1905 paper on the light quanta, Einstein had also accounted for old evidence. He explained the photoelectric effect observed by Philipp Lenard in 1900. In Section $\S8$ of his 1905 paper on the light quanta, Einstein treats the photoelectric effect. He is guided by the following principle: "the energy of light is distributed discontinuously in space” and writes: "The usual conception, that the energy of light is continuously distributed over the space through which it travels, meets with especially great difficulties when one attempts to explain the photoelectric phenomena, which are described in the pioneering work by Mr. Lenard" (\cite{Einstein}, p. 133, p. 145). Einstein had read Lenard's paper on the photoelectric effect four years earlier in 1901 (\cite{Lenard}; Einstein to Marić, probably May 28, 1901, \cite{CPAE1}, Doc. 111). But in his 1905 paper on the light quanta, he insists that his account of the photoelectric effect is deduced by his light quanta principle and not induced from Lenard's experimental result. He concludes his exposition saying: "As far as I can see, our conception does not contradict with the properties of the photoelectric effect observed by Mr. Lenard" (Einstein, 1905c, p. 145, p. 147). Hempel explains that by means of a given hypothesis, we can deduce special predictions, which have the form of observation laws (\cite{Hempel}, p. 101). On the Nobel Prize website it is written: "The Committee says that the most important application of Einstein’s photoelectric law and also its most convincing confirmation has come from the use Bohr made of it in his theory of atoms, which explains a vast amount of spectroscopic data" (\cite{Ekspong}). It therefore seems that in the case of the photoelectric effect: $P_t(e) \neq 1$. 

On the other hand, as discussed above, Einstein's general relativity confirms the perihelion of Mercury and therefore, to the Bayesian the prior is: $P_t(e) = 1$. 
More specifically, Einstein was already interested in the problem of Mercury's perihelion from early on in his search for a new relativistic theory of gravitation. He wrote to his close friend Conrad Habicht in 1907 that he hoped his gravitation theory would explain the anomalous advance of Mercury's perihelion (Einstein to Habicht, December 24, 1907, \cite{CPAE5}, Doc. 69). Accordingly, Einstein had long known that the anomalous advance of Mercury's perihelion could not be explained by Newtonian theory and by 1913, he also knew the value for the anomalous secular advance of Mercury's perihelion. 
Hence, evidence $e$ was known to Einstein in 1907, before the \emph{Entwurf} theory was introduced by him in 1913. In $t=1915$, Einstein presented his general theory of relativity $T$. Because $e$ was already known at $t=1915$, for Einstein $P_t(e) = 1$. Further, because $P_t(e) = 1$, the likelihood of $e$ given $T$, $P_t(e, T)$, is also $1$. We then have:

\begin{equation} \label{equation 3}
P_t(T, e) = \frac{P_t(T) P_t(e, T)}{P_t(e)} =  P_t(T). \vspace{1mm} 
\end{equation}

\noindent The posterior probability $P_t(T, e)$ is therefore the same as the prior probability $P_t(T)$ and $e$ cannot constitute evidence for Einstein's $T$ and “we have the absurdity that old evidence cannot confirm new theory”, says Glymour (\cite{Glymour1}, pp. 85-86).

Many philosophers attempted to solve and dissolve the old evidence problem (see \cite{Earman}, p. 120 and \cite{Garber}, to name only two).  

Bas van Fraassen has responded to Glymour's claim saying that although the measurements of the anomalous advance of Mercury’s perihelion had long been made, when Einstein showed in 1915 that the above measurements do, indeed, confirm his new theory, \emph{we} updated our credence (\cite{Fraassen}, p. 154):

\begin{quote}
Indeed, when the scientific community got to the point when it could say that the advance in the perihelion of Mercury confirmed Einstein's theory, they had indeed just newly learned \emph{something} -- though not \emph{that} fact about
Mercury, but another fact that has to do with it -- had conditionalized on that something, and the conditionalization had increased their credence in Einstein's theory.    
\end{quote}
\vspace{1mm} 

This solution is more reasonable. But it seems to me that the problem in Bayesian confirmation theory persists.  

\section{Why did Einstein not invite Besso as a co-author in 1915?} \label{3}

Besso collaborated with Einstein on a project (the \emph{Einstein-Besso manuscript}) that contains a few methods and equations that are similar to the ones one finds in Einstein's November 18, 1915 paper on the perihelion motion of Mercury. Some time in November 1915, Einstein realized that unlike the 1913 calculations, he could now demonstrate that the $43^{''}$ does, indeed, confirm his new general theory of relativity. Unlike the \emph{Einstein-Besso manuscript}, with Einstein's new theory and the Christoffel symbols, it seemed like everything simply fell into place.     

With his new 1915 general theory of relativity, Einstein quickly achieved the correct advance of Mercury's perihelion in the field of a static Sun by applying the methods he had already worked out two years earlier with Besso. 
Recall that Einstein wrote to Hilbert on the very day he submitted (to the \emph{Proceedings of the Royal Prussian Academy of Sciences, Berlin}) the November 18, 1915 paper on the perihelion motion of Mercury (see section \ref{2}). Hilbert wrote back the next day saying: "If I could calculate as fast as you can" (Hilbert to Einstein, November 19, 1915, \cite{CPAE8}, Doc. 149). 
According to Janssen and Renn, Einstein did not bother to tell Hilbert that he had done very similar calculations before in the \emph{Einstein-Besso manuscript}. "He probably enjoyed giving Hilbert a taste of his own medicine". It is only because the \emph{Einstein-Besso manuscript} "ended up in Besso’s hands rather than Einstein’s, who almost certainly would have discarded it, that we can belatedly call Einstein’s bluff trying to put one over on Hilbert", say Janssen and Renn (\cite{Janssen}, pp. 5-6).   

Hence, Einstein did not acknowledge his earlier work with Besso. Neither did he mention Besso's name in his 1915 paper explaining the anomalous precession of Mercury.  
"Besso had always been thought of as an important sounding board for Einstein, never as a serious scientific
collaborator", say Janssen and Renn (\cite{Janssen}, p. 6). When the \emph{Einstein-Besso manuscript} was discovered, scholars realized that Einstein had already calculated the perihelion motion of Mercury in 1913 in collaboration with Besso.
Janssen and Renn then write about the November 18, 1915 paper (\cite{Janssen}, p. 7):

\begin{quote}

Still, given the undeniable importance of his earlier
calculations, one can legitimately ask why Einstein did not invite Besso as a co-author. As it happened, Besso did not even get an acknowledgment. We suspect that this is largely because of the race Einstein perceived himself to
be in with Hilbert. He was in a hurry and it probably never even occurred to him to ask Besso to write a joint paper on the topic.
\end{quote}
\vspace{1mm} 
Janssen and Renn add that "An unusually high number of typos in the paper suggests that it was written in haste".
\vspace{1mm} 

There seem to me to be two main reasons why Einstein did not invite Besso as a co-author in 1915:

\vspace{1mm} 
\noindent 1) \emph{First}, it appears that Einstein still considered Besso as a sounding board, even though Besso undertook calculations with Einstein in the \emph{Einstein-Besso manuscript}. When Einstein wrote a series of letters to Besso between 1913 and 1916 (\cite{Besso}), and described to him, step by step, his discoveries of general relativity, Besso once again functioned as the committed sounding board.
On the other hand, from 1912, apparently Einstein considered Grossmann, and not Besso, as his partner and collaborator while Besso remained Einstein's closest friend and sounding board. 

Let us pause on this point. 
Janssen and Renn write (\cite{Janssen}, pp. 7-8, p. 27): "Besso nonetheless took responsibility for key parts of their joint project and corrected some errors in Einstein’s
calculations". Besso corrected Einstein’s overly crude approximations of Mercury's precession of Mercury’s perihelion
produced by the field of the Sun. Recall that on page 28 of the \emph{Einstein-Besso manuscript}, Einstein recorded the value for the “precession in 100 years” which was too large by a factor of 100, i.e. $1821^{"}$. On page 35 of the \emph{Einstein-Besso manuscript}, Besso located the source of
the error in the value Einstein reported on page 28. On page 30 Einstein made the same correction (see discussion in section \ref{1}). Janssen and Renn claim that "The exclamation point on [p. 35] suggests that Besso found the mistake first". 

It was not the first time that Einstein made errors in calculations. Recall that a few pages ahead, on pages 40-41 of the \emph{Einstein-Besso manuscript}, one can see that Einstein checked whether the rotation metric was a vacuum solution of the field equations he established. He did a sign error and a few mistakes and came to the belief that the rotation metric was a solution of the new \emph{Entwurf} field equations (\cite{CPAE4}).
Recall from section \ref{1} that according to Janssen and Renn, in August 1913, Besso told Einstein that the rotation metric was not a solution of the \emph{Entwurf} field equations. \emph{But at the end of the day, before November 1915, Einstein stubbornly clung to his \emph{Entwurf} field equations like a butterfly to a flower}. 

Think of the following example. Much later, in 1916, Einstein published a paper on gravitational waves in which he presented a solution that did not satisfy the unimodular coordinates. In obtaining the above solution, he made a mathematical error. This error caused him to obtain three different types of waves compatible with his approximate field equations: longitudinal waves, transverse waves and a new type of wave. He added an Addendum in which he suggested a solution to this problem: A comeback of the restrictive condition. He suggested that for systems in unimodular coordinates waves of the third type transport energy. Einstein's colleagues (Willem de Sitter, Tulio Levi-Civita, Erwin Schrödinger, and Gunnar Nordström) demonstrated to him that for systems in unimodular coordinates the gravitational wave of a mass point carry no energy, but Einstein tried to persuade them that he had actually not made a mistake in his gravitational waves paper.\footnote{After 1916 Nordström gave up his scalar theory of gravitation and started to work on Einstein's new general theory of relativity.} The stubborn Einstein, however, eventually accepted his colleagues' results and dropped his restrictive condition. This finally led him to discover plane transverse gravitational waves. Einstein confided in Gustav Mie and told him that his 1916 gravitational wave paper contained calculation errors, and that he intended to present a cleaned-up version of the paper to the Prussian Academy (Einstein to Gustav Mie, December 29, 1917, \cite{CPAE8}, Doc. 421). Einstein was the sole author of the new paper published in 1918. Neither de Sitter, nor Levi-Civita, Schrödinger, and Nordström were co-authors of the new 1918 paper on gravitational waves, even though they corrected Einstein's errors and calculations from the 1916 paper. 

It seems that the reason why Einstein made errors in calculations is his ability to think in images, thought experiments, and entertain non-verbal images. Einstein first used images (anchored in the unconscious) to solve problems in science; he first imagined thought experiments. Words, as they are written or spoken, came later, when he expressed his discoveries through the language of logical connections and the discovery of universal formal principles (see \cite{Weinstein}, chapter E). Einstein's sister Maja reported that in the Gymnasium he was supposed to begin the study of algebra and geometry at the age of thirteen. By that time however he already had a predilection for solving complicated problems in applied arithmetic, although the computational errors he made kept him from appearing particularly talented in the eyes of his teachers (\cite{CPAE1}, pp. 1xi, pp. xx). Thus, even as a child Einstein would make computational errors. 
\vspace{1mm} 

\noindent 2) \emph{Secondly}, and most importantly, in the November 1915 perihelion motion of Mercury paper, Einstein wrote the so-called "area law" (angular momentum conservation) and the law of conservation of energy for the orbit of Mercury in polar coordinates. In the \emph{Einstein-Besso manuscript}, Besso had already written the "area law" - and defined the area constant and the area speed - and wrote the equation for the law of energy conservation for the orbit of Mercury (\cite{CPAE4}, p. 8).  
However, in the \emph{Einstein-Besso manuscript}, Besso copied equations directly from Einstein's 1913 \emph{Entwurf} paper. In the \emph{Entwurf} paper, Einstein obtained the energy and momentum of the mass point, respectively. Besso copied the equations of motion for the mass point, the energy and the momentum of the mass point, and the Lagrangian directly from Einstein's \emph{Entwurf} paper and derived the angular momentum conservation, which he called the "area law" (\cite{Grossmann}, p. 7). 
But according to Besso's angular momentum conservation expression (the "area law"), Kepler's second law did not hold for the \emph{Entwurf} theory. There was, therefore, no warrant for calling it an "area law".

In 1915, Einstein proceeded by using the "area law" and the law of conservation of energy for the orbit of Mercury, the components of the gravitational field to the first-order approximation and the components of the gravitational field to the second-order approximation. Subsequently, using the equations of the geodesic line in terms of the Christoffel symbols, Einstein obtained the equation of motion of Mercury to the second-order approximation. Eventually, from Einstein's derivation it followed that Kepler's laws hold exactly. 

Accordingly, Einstein need not have invited Besso as a co-author of his 1915 paper. 
Einstein ends his 1905 relativity paper by noting that when he worked on the problem discussed in the relativity paper, Besso faithfully stood by him. Einstein is indebted to him for several "valuable suggestions" (\cite{Einstein4}, p. 921). He does not end the November, 18, 1915 perihelion motion of Mercury paper with a similar acknowledgment to Besso.    
Einstein though probably felt guilty about not \emph{thanking} Besso in his November 18, 1915 paper, because much later, he told his assistant Leopold Infeld that he should look up the literature to quote previous scientists: "Do it by all means. Already I have sinned too often in this respect" (\cite{Infeld}, p. 277). 
\vspace{5mm} 

\noindent This work is supported by ERC advanced grant number 834735.

\end{document}